\documentclass[a4paper,11pt]{article}
\usepackage{amsfonts}
\usepackage{amsmath}
\usepackage{amssymb}
\usepackage{graphicx}
\usepackage{array}
\usepackage{booktabs}
\usepackage{multirow}
\usepackage{makecell}
\usepackage{float}
\pdfoutput=1 

\usepackage{mymacros}
\usepackage{chngcntr}
\usepackage{verbatim}
\usepackage{amsthm}
\usepackage{graphics}
\usepackage{xcolor}
\usepackage{mathrsfs}
\usepackage{bbold}
\usepackage{cases}
\usepackage{epsfig}
\usepackage{epstopdf}
\usepackage{hyperref}
\usepackage{amsfonts}
\usepackage{hyperref}
\usepackage{jheppub} 

\usepackage[T1]{fontenc} 

\hypersetup{
	colorlinks=true,
	linkcolor=blue,
	filecolor=blue,
	urlcolor=blue,
	citecolor=blue,
}

\title{\boldmath Motions of spinning particles and chaos bound in Reissner-Nordstr\"om spacetime}

\author[a]{Chuang Yang, \footnote{E-mail: chuangyangyc@hotmail.com}}
\author[a,b]{Deyou Chen, \footnote{E-mail: deyouchen@hotmail.com }}
\author[a]{Yongtao Liu, \footnote{E-mail: lytao@foxmail.com}}

\affiliation{$^{a}$School of Science, Xihua University, Chengdu 610039, China}
\affiliation{$^{b}$College of Physics, Chengdu University of Technology, Chengdu 610059, China}

\abstract{Previous research showed that the chaos bound proposed in \cite{MSS} can be violated under specific conditions within the scalar fields surrounding black holes. In this paper, we explore motions of spinning particles orbiting a Reissner-Nordstr\"om black hole and examine whether this bound is violated in the spinor field of this black hole. For the neutral particle, when its spin magnitude surpasses a specific threshold, the value of the exponent exceeds the surface gravity, resulting in a violation of the bound. Given a fixed total angular momentum of the particle, when its spin direction is anti-aligned with the angular momentum direction, the exponent value is greater than that when the two directions are aligned. For the charged particle, taking into account the influence of the electromagnetic force, we find that for relatively large angular momenta, although the electromagnetic force does not change the trend of the exponent's variation with respect to spin and angular momentum, and only modifies its values, it still leads to the violation. Therefore, the chaos bound violations are observed in the spinor field.}

\begin{document} 
	\maketitle
	\flushbottom
	
\section{Introduction}

In the recent work, Maldacena, Shenker and Stanford proposed a significant conjecture \cite{MSS}: the Lyapunov exponent(LE) in thermal quantum systems with a large number of degrees of freedom is subject to a temperature-dependent upper bound, known as the chaos bound, expressed as

\begin{eqnarray}
\lambda \leq \frac{2\pi T}{\hbar}.
\label{eq1.1}
\end{eqnarray}

\noindent where $\lambda$ denotes the exponent, and $T$ is the system temperature. This bound was identified via the AdS/CFT correspondence \cite{JMM}, by conducting thought experiments concerning shock waves in the vicinity of the black hole(BH) horizons \cite{SS1}. This conjecture has since garnered substantial attention and received extensive support from subsequent research. In Einstein gravity, the exponent associated with a BH is precisely given by $2\pi/ \beta$, with $\beta$ denoting the inverse temperature \cite{SS2,SS3}. This finding indicates that the bound is saturated within this gravitational system. The saturated bound can provide sufficient conditions for the existence of Einstein gravity dual. Another notable example of supporting evidence is the saturated chaos bound that has been obtained in the Sachdev-Ye-Kitaev(SYK) model \cite{SYK1,SYK2,PR,MS,CFCZZ}. Furthermore, the investigation into the effects of electromagnetic and scalar forces on the exponent for classical chaos of particles outside the BH revealed that its value remains independent of the types of external forces and particle mass, obeying the inequality $\lambda \leq \kappa$, where $\kappa$ is the surface gravity of the BH \cite{HT}. Given the relationship between the surface gravity and temperature, this classical result aligns with findings in the quantum thermal systems, thereby indicating a universal manifestation of the bound across both classical and quantum systems.

Despite widespread support for the conjecture of Maldacena, Shenker and Stanford, there are some cases which indicate that this bound can be violated \cite{HT1,HT2,HT3,HT4,HT5,PZDA1,PZDA2}. In the AdS spacetime, the analyses of closed string dynamics indicate that the exponent can be expressed as $\lambda = 2n\pi T$,  where $n$ is a string winding number \cite{HT1}, which provides a modification to the bound. Subsequently, Hashimoto and Sugiura developed a reparametrization-independent method for analytically estimating the exponent \cite{HT2}. They found that the causality gives a universal upper bound, namely $\lambda \propto E$ $ (E \rightarrow \infty)$, and the bound is violated in the particular potentials. Related work further demonstrated that the BTZ BH exhibits two distinct exponents: one that adheres to the bound and another that exceeds it \cite{HT3,HT4}. Nevertheless, it is possible to prevent the occurrence of the latter by defining the effective temperatures of the left and right moving modes \cite{HT5}.

In classical systems, when charged particles around a BH are subjected to electromagnetic forces, they may reach equilibrium states. By adjusting the charge-to-mass ratios of the particles, one can make them move close to the event horizon. Based on this, the studies have been conducted on the exponent of scalar particle chaos in the vicinity of the event horizons. It was found that the chaos bound violations occur in multiple spherically symmetric spacetimes \cite{ZLL}, as well as Kerr-Newman \cite{KG1} and Kerr-Newman-AdS spacetimes \cite{KG2}. However, the bound for the particles in Reissner-Nordstr\"om (RN) and RN-AdS spacetimes remains unviolated. The aforementioned LEs are obtained through the calculations of the effective Lagrangians and effective potentials of the particles. In fact, the exponents can also be derived by computing eigenvalues of a Jacobian matrix. By utilizing the latter approach and exploring the influence of the particle's angular momentum on the exponents, it was observed that the violations of the bound also manifest around the RN, RN-AdS \cite{LG1} and Kiselev BHs \cite{GCYW1}. Meanwhile, the spatial regions where the violation occurs were also identified \cite{GCYW2}. Other research concerning the chaos bound are referred to \cite{LG2,LTW1,LTW2,LTW3,LTW4,LTW5,LTW6,LTW7,LTW8,LTW9,LTW10}. All these research endeavors have unveiled the phenomenon of chaos bound violation for scalar fields in the classical systems. Consequently, the question arises as to whether a similar violation also exists for spinor fields.

In this paper, we explore motions of spinning particles orbiting an RN BH, and examine whether the values of LEs characterizing their chaotic motion within the spinor field of this BH surpass the chaos bound. The impacts of the electromagnetic force and orbital angular momentum on the exponents for the particles outside this BH have already been studied, but these results were all obtained in the scalar field. To study the influences on the exponents of chaos for the spinning particles, we first employ the approach proposed in \cite{ZHA} to solve the particles' radial equations of motion. The exponents are obtained through calculations of the effective potentials. Then, we discuss the influences of the spin and total angular momentum of a neutral particle on the exponent and test the bound. When the particle is charged, we examine the impacts of factors such as the electromagnetic force and spin of the particle on the exponent. In our work, the same spin magnitude can affect the values of the exponents depending on whether the spin direction of the particle is aligned or anti-aligned with the $z$-axis direction, thereby influencing the threshold values of relevant parameters when the bound is violated. Therefore, both cases should be considered. The result shows that the bound is violated in this spinor field.

The rest of this paper is organized as follows. In Section \ref{sec2}, the motions of the spinning particles around the RN BH are explored, ultimately resulting in the derivation of the particles' radial equations. In Section \ref{sec3}, we first investigate the influences of the spin and total angular momentum of the neutral particle on the LE and examine the chaos bound. Subsequently, we take into account the scenario where the particle is charged. Combining the electromagnetic force acting on the particle with the spin and total angular momentum, we further study the exponent and examine the bound. The final section presents our conclusions and discussions.

\section{Motions of spinning particles around RN BH}\label{sec2}

\subsection{Review of MPD equations}\label{sec2.1}

This section offers a concise overview of the dynamics associated with a charged spinning particle traversing a gravitational field. The motion of this particle is governed by the Mathisson–Papapetrou–Dixon (MPD) equations \cite{HH}

\begin{align}
\frac{D p^{\mu}}{D \tau} &= -\frac{1}{2}R^{\mu}_{\nu\alpha\beta}u^{\nu}S^{\alpha\beta} - mqF^{\mu}_{\nu}u^{\nu}, \label{eq2.1}\\
\frac{D S^{\mu\nu}}{D \tau} &=  p^{\mu} u^{\nu} - u^{\mu} p^{\nu},
\label{eq2.2}
\end{align}

\noindent where $\frac{D}{D\tau}$ represents the covariant derivative along the particle’s trajectory, with $\tau$ being an affine parameter. The four-momentum and four-velocity vectors of the particle are denoted by$p^{\mu}$ and $u^{\mu}=\frac{dx^{\mu}}{d\tau}$, respectively. The mass and charge of the particle are symbolized by $m$ and $q$. The Riemannian tensor is given by $R^{\mu}_{\nu\alpha\beta}$, and $S_{\mu\nu}$ stands for an antisymmetric spin tensor. Eq. (\ref{eq2.2}) illustrates the interaction between the spin and orbital motion of the particle. Additionally, $F_{\mu\nu} = A_{\mu,\nu} - A_{\nu,\mu}$ is an electromagnetic field tensor of the spacetime. In \cite{DR1,DR2}, these equations were modified in ultra relativistic limit.

In order to precisely establish the relationship between the four-velocity and four-momentum of the particle, and to avoid any infringement upon the timelike characteristic of the four-velocity, the Tulczyjew-Dixon spin supplemental condition(TDSSC),

\begin{eqnarray}
S^{\mu\nu}p_{\nu} = 0,
\label{eq2.5}
\end{eqnarray}

\noindent is required \cite{WT1,WT2}. These two physical quantities, $S^{\mu\nu}$ and $p_{\mu}$, are respectively characterized by the conserved quantities of mass $m$ and of spin magnitude $\tilde{S}$, as follows

\begin{align}
-m^2 &= p^{\mu} p_{\mu},  \label{eq2.3} \\
2\tilde{S}^2 &= S^{\mu\nu}S_{\mu\nu}.
\label{eq2.4}
\end{align}

\noindent The unit vector parallel to the four-momentum is defined as $v^{\mu} = \frac{p^{\mu}}{m}$ and its relation with the four-velocity vectors is given by \cite{HZ1,HZ2}

\begin{eqnarray}
u^{\mu} - v^{\mu} = \frac{2S^{\mu\nu}v^{\alpha}\left(R_{\nu\alpha\beta\gamma} S^{\beta\gamma}+2qF_{\nu\alpha}\right)}{ S^{\nu\alpha}\left(R_{\nu\alpha\beta\gamma} S^{\beta\gamma}+2qF_{\nu\alpha}\right)+4m^2} .
\label{eq2.6}
\end{eqnarray}

\noindent It is evident that the four-velocity and the four-momentum are not parallel owing to the spin. 

\subsection{Motions of spinning particles}\label{sec2.2}

We investigate the motion of a charged spinning particle orbiting the RN BH within its equatorial plane. This BH is a spherically symmetric solution within the Einstein gravity coupled to the Maxwell field, and its metric is given by \cite{ZLL}

\begin{eqnarray}
ds^2 = -f(r)dt^2 + \frac{1}{f(r)}dr^2 + r^2\left(d\theta^2 + \sin^2\theta d\phi^2\right),
\label{eq3.1}
\end{eqnarray}

\noindent with an electromagnetic potential $A_t = \frac{Q}{r}$, where

\begin{eqnarray}
f(r) = 1-\frac{2M}{r} +\frac{Q^2}{r^2},
\label{eq3.2}
\end{eqnarray}

\noindent $M$ and $Q$ denote the mass and charge of the BH, respectively. When $Q=0$, the metric is reduced to the Schwarzschild metric. For the equation $f(r)=0$, there exist two roots, namely $r_{\pm} = M \pm\sqrt{M^2-Q^2}$. These roots correspond to the event horizon $r_+$ and the inner horizon $r_-$. The surface gravity is

\begin{eqnarray}
\kappa = \frac{r_+ -r_-}{2r_+^2}.
\label{eq3.2.0}
\end{eqnarray}

\noindent To solve the equations of motion for the particle under consideration, we employ the methodology delineated in \cite{ZHA}. In that work, the equations governing the motion of neutral particles within the spherically symmetric spacetimes have been successfully resolved. Other studies on the motions for spinning particle are referred to \cite{BF1,BF2,BF3}.

In the given spacetime, there exist two Killing vectors, specifically $\left(\frac{\partial}{\partial t}\right)^a$ and $\left(\frac{\partial}{\partial \phi}\right)^a$. The first Killing vector is associated with a conserved quantity, namely the energy $E$, which is expressed as

\begin{eqnarray}
E= f p^t - \frac{1}{2}f^{\prime}S^{tr} + mqA_t.
\label{eq3.3}
\end{eqnarray}

\noindent The Killing vector $\left(\frac{\partial}{\partial \phi}\right)^a$ is associated with the angular momentum oriented along the $z$-axis,

\begin{eqnarray}
L=r^2p^{\phi} + rS^{r\phi}.
\label{eq3.4}
\end{eqnarray}

\noindent This angular momentum arises from the coupling interaction between the spin angular momentum and orbital angular momentum of the particle. When it takes on a positive value, it indicates a directional alignment along the $z$-axis. In contrast, a negative value implies a directional orientation opposite to the $z$-axis. When the particle moves in the equatorial plane, the expressions for the mass and the magnitude of the spin are given by

\begin{align}
m^2 &=(p^t)^2f -\frac{(p^r)^2}{f} - r^2(p^{\phi})^2, \label{eq3.5}\\
\tilde{S}^2 &= -(S^{tr})^2 -fr^2(S^{t\phi})^2 +\frac{r^2(S^{r\phi})^2}{f}.
\label{eq3.6}
\end{align}

\noindent In the RN spacetime, the TDSSC takes on the following specific form

\begin{align}
\frac{p^rS^{tr}}{f} + r^2 p^{\phi}S^{t\phi} &= 0, \label{eq3.7}\\
 r^2 p^{\phi}S^{r\phi} + fp^tS^{tr} &= 0, \label{eq3.8}\\
\frac{1}{f}p^{r}S^{r\phi} - fp^t S^{t\phi} &= 0.
\label{eq3.9}
\end{align}

\noindent The momentum equations (\ref{eq2.1}) corresponding to this metric are explicitly formulated as

\begin{align}
\dot{p}^t +\frac{p^r f^{\prime}}{2f}+\frac{\dot{r}p^t f^{\prime}}{2f} -\frac{r\dot{\phi}S^{t\phi} f^{\prime}}{2} -\frac{\dot{r}S^{tr} f^{\prime\prime}}{2f} +\frac{m\dot{r}qA_{t,r}}{f} &= 0, \label{eq3.10}\\
\dot{p}^r +\frac{p^t ff^{\prime}}{2} - \dot{\phi}p^{\phi} fr -\frac{r\dot{\phi}S^{r\phi} f^{\prime}}{2}-\frac{S^{tr} ff^{\prime\prime}}{2} -\frac{\dot{r}p^{r} f^{\prime}}{2f} +\frac{\dot{r}qA_{t,r}}{f} +mqA_{t,r}f &= 0, \label{eq3.11}\\
\dot{p}^{\phi} +\frac{\dot{\phi}p^r }{r}+ \frac{\dot{r}p^{\phi} }{r} -\frac{\dot{r}S^{r\phi} f^{\prime}}{2fr} -\frac{S^{t\phi} ff^{\prime}}{2r} &= 0.
\label{eq3.12}
\end{align}

\noindent In the above equation, the over-dot sign indicates a derivative with respect to the coordinate time $t$, while the prime symbol denotes a derivative with respect to the radial coordinate $r$. The spin equations, derived from Eq. (\ref{eq2.2}) through mathematical procedures, are presented as follows

\begin{align}
\dot{S}^{tr} +p^r -\dot{r}p^t - fr\dot{\phi}S^{t\phi}  &= 0, \label{eq3.13}\\
\dot{S}^{t\phi} -\dot{\phi}p^t +p^{\phi} + \frac{\dot{\phi}S^{tr}+\dot{r}S^{t\phi}}{r} + \frac{S^{r\phi}f^{\prime}+\dot{r}S^{t\phi}f^{\prime}}{2f}  &= 0, \label{eq3.14}\\
\dot{S}^{r\phi} -\dot{\phi}p^r +\dot{r}p^{\phi} + \frac{\dot{r}S^{r\phi}}{r} + \frac{S^{t\phi}ff^{\prime}}{2} - \frac{\dot{r}S^{r\phi}f^{\prime}}{2f}  &= 0.
\label{eq3.15}
\end{align}

\noindent Using Eqs. (\ref{eq3.5})-(\ref{eq3.8}), we get 

\begin{align}
S^{tr} = \pm\frac{Srp^{\phi}}{m}.
\label{eq3.16}
\end{align}

\noindent By employing the aforementioned relation in conjunction with Eqs. (\ref{eq3.3}), (\ref{eq3.4}) and (\ref{eq3.8}), we obtain

\begin{align}
p^{\phi} =\frac{m^2L\mp m(E-qA_t)\tilde{S}}{m^2r^2 - \frac{1}{2}r\tilde{S}^2f^{\prime}},
\label{eq3.17}
\end{align}

\noindent where the "$\pm$"  signs indicate the relative direction between the spin and the $z$-axis. Specifically, a "+" sign denotes that the two directions are aligned, while a "$-$" sign indicates an anti-aligned direction between them. To get $p^t$, it is necessary to solve Eqs. (\ref{eq3.3}), (\ref{eq3.16}) and (\ref{eq3.17}). After a series of mathematical manipulations, the solution is found to be

\begin{align}
p^t = \frac{2m^2r(E-qA_t)\mp m\tilde{S}Lf^{\prime}}{(\tilde{S}^2f^{\prime} - 2m^2r )f}. 
\label{eq3.18}
\end{align}

\noindent Subsequently, by substituting Eqs. (\ref{eq3.17}) and (\ref{eq3.18}) into Eq. (\ref{eq2.3}), we can rigorously derive the expression for the conjugate momentum in the radial direction, which is given by 

\begin{align}
p^r= \pm\sqrt{p_t^2-f\left(m^2+\frac{p_{\phi}^2}{r^2}\right)}.
\label{eq3.19}
\end{align}

\noindent In the above equation, $p^2_t=(f p^t)^2$ and $p^2_{\phi}= (p^{\phi}r^{2})^2$. The symbol "$\pm$" serves to indicate the direction of the particle's radial motion. More precisely, the "$+$" sign corresponds to an outward radial motion of the particle, while the "$-$" sign represents an inward one. The effective potential of the particle in the radial direction is of critical importance for determining the LE. To achieve this, it is essential to derive the radial equation with respect to the coordinate time. Initially, by employing the constraints given by Eqs. (\ref{eq3.7}) and (\ref{eq3.8}) in conjunction with Eq. (\ref{eq3.16}), we analytically determine the components of the spin tensor 

\begin{align}
S^{t\phi}= \mp \frac{\tilde{S}p^r}{mrf}, \quad S^{r\phi}= \mp \frac{p^t \tilde{S}f}{mr}.
\label{eq3.20}
\end{align}

\noindent Subsequently, through an analysis of Eqs. (\ref{eq3.9}), (\ref{eq3.13}) and (\ref{eq3.20}), we obtain

\begin{align}
\frac{(p^r -\dot{r}p^t)\tilde{S}^2f^{\prime}}{2m^2r} \pm \frac{\tilde{S}(r\dot{p}^{\phi}+ \dot{r}p^{\phi})}{m} - p^r + \dot{r}p^t -\dot{S}^{tr} = 0.
\label{eq3.21}
\end{align}

\noindent To derive the time derivative of the radial coordinate $\dot{r}$, we utilize the aforementioned equation in combination with Eq. (\ref{eq3.16}). Through a series of mathematical manipulations, we get the expression

\begin{align}
\dot{r}= \frac{p^r}{p^t}. 
\label{eq3.22}
\end{align}

\noindent The equation governing the motion in the $\phi$-direction can be derived through the solution of the equations presented earlier. Nevertheless, given that our research emphasis is placed on the equation of radial motion, owing to its close connection with the effective potential to be elaborated upon in subsequent sections, the explicit form of the $\phi$-direction motion equation is not presented here.

\section{LE and chaos bound}\label{sec3}

It has been discovered that when the charged scalar particle moves chaotically around the RN BH, its LE can exceed the surface gravity, which implies that the chaos bound would be violated \cite{LG1}. We now examine whether the motions of the spinning particles around this BH cause the violation of the bound. In the calculations, we set $M=m=1$. In the preceding section, $\tilde{S}$ represents the spin magnitude, and the "$\pm$" signs are incorporated into Eqs. (\ref{eq3.17}) and (\ref{eq3.18}). To facilitate analysis, we herein adopt the parameterization $S= \pm\tilde{S}$ in this section, thereby allowing for a systematic investigation into two distinct physical configurations: (1) the scenario where the spin direction aligns with the $z$-axis, with the positive sign being adopted, and (2) the scenario where the spin direction anti-aligns with to the $z$-axis, with the negative sign being taken.

\subsection{LE}\label{sec3.1}

The effective potential of the particle serves as a crucial analytical tool for investigating its motion. It not only determines the stability of orbital trajectories but also maintains a close relationship with the LE. In \cite{CMBWZ}, the authors discovered, through exploring the relationships among unstable null geodesics, the exponent and quasinormal modes, that the exponent is correlated with the second derivative of the potential for the radial motion. Here, we take into account a small perturbation around the local maximum of the potential to calculate the exponent. The radial equation of motion in terms of the coordinate time is given by

\begin{align}
\frac{1}{2}m\dot{r}^2 +\mathcal{V}_{eff}=0. 
\label{eq3.23}
\end{align}

\noindent where $\mathcal{V}_{eff}$ is defined as the effective potential. By comparing the aforementioned equation with Eq. (\ref{eq3.22}), it is found that the expression for the potential is  

\begin{align}
\mathcal{V}_{eff} = -\frac{m}{2} \left(\frac{p^r}{p^t}\right)^2. 
\label{eq3.23-1}
\end{align}

\noindent Given that the motion of the particle near an unstable equilibrium orbit is considered and $r_0$ is designated as the location of this orbit, the radial position of the particle can be expressed as $r(t)= r_0 +\epsilon(t)$. We subsequently substitute this expression into Eq. (\ref{eq3.23}), perform Taylor expansions on it, and thereby obtain

\begin{eqnarray}
\frac{1}{2}\left(m\dot{\epsilon}^2 + \mathcal{V}^{\prime\prime}_{eff}(r_0) \epsilon^2\right) + \mathcal{V}_{eff}(r_0)+ \mathcal{O}(\epsilon) \simeq 0, 
\label{eq3.24}
\end{eqnarray}

\noindent where $\epsilon = \epsilon(t)$ and $\mathcal{O}(\epsilon)$ represents higher order terms with respect to $\epsilon$. To simplify calculations, we let $\mathcal{V}_{eff}(r_0)=0$ \cite{CMBWZ}. By neglecting the higher order terms, we obtain the solution of the aforementioned equation in the following form: $\epsilon = c_0e^{\pm \lambda t}$, where $c_0$ is an integral constant, and the positive sign is adopted in the "$\pm$" notation. If $\lambda$ is a real number, an exponential growth of the trajectory deviation with respect to the time has the potential to give rise to the onset of chaos. Here, $\lambda$ is defined as a LE, and its relationship with the potential is gotten as \cite{LTW2,CHL}

\begin{eqnarray}
\lambda^2 =-\frac{\mathcal{V}^{\prime\prime}_{eff}(r_0)}{m}.
\label{eq3.25}
\end{eqnarray}

\noindent By employing Eq. (\ref{eq3.23-1}), the exponent is rewritten as follows

\begin{align}
\lambda^2 = \left.\frac{1}{2} \frac{d^2}{dr^2}\left(\frac{p^r}{p^t}\right)^2  \right |_{r=r_0}. 
\label{eq3.26}
\end{align}

\noindent A positive exponent implies the divergence of the radial motion of the particle, while a negative one suggests its convergence. The exponent that characterizes the chaotic motion of a scalar particle is recovered by ordering $S=0$.

\subsection{Neutral particle}\label{sec3.2}

The LE for the chaotic motion of the neutral particle on an unstable equilibrium orbit is first calculated. In the previous subsection, the effective potential was set to zero to solve the exponent, and this value is retained in the present analysis. The position of the unstable equilibrium orbit is determined by the conditions $V_{eff}^{\prime}(r_0) =0$ and $V_{eff}^{\prime\prime}(r_0) <0$. By utilizing Eqs. (\ref{eq3.2}), (\ref{eq3.2.0}), (\ref{eq3.17}), (\ref{eq3.18}), (\ref{eq3.19}) and (\ref{eq3.26}), we numerically calculate the exponent and the surface gravity. Subsequently, we generate plots, with the aim to elucidate the influences of the particle's spin and total angular momentum, as well as the BH's charge, on the exponent. 

\begin{figure}[h]
	\centering
	\begin{minipage}[t]{0.49\textwidth}
		\centering
		\includegraphics[width=7cm,height=6cm]{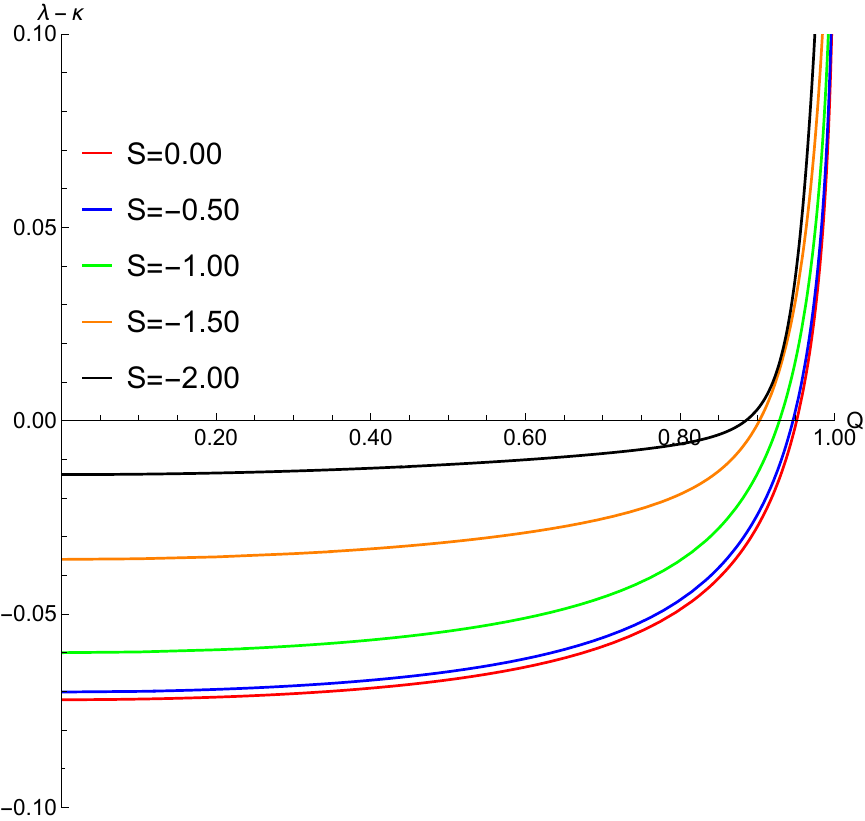}
		\subcaption{}
		\label{4f2-a}
	\end{minipage}
	\begin{minipage}[t]{0.49\textwidth}
		\centering
		\includegraphics[width=7cm,height=6cm]{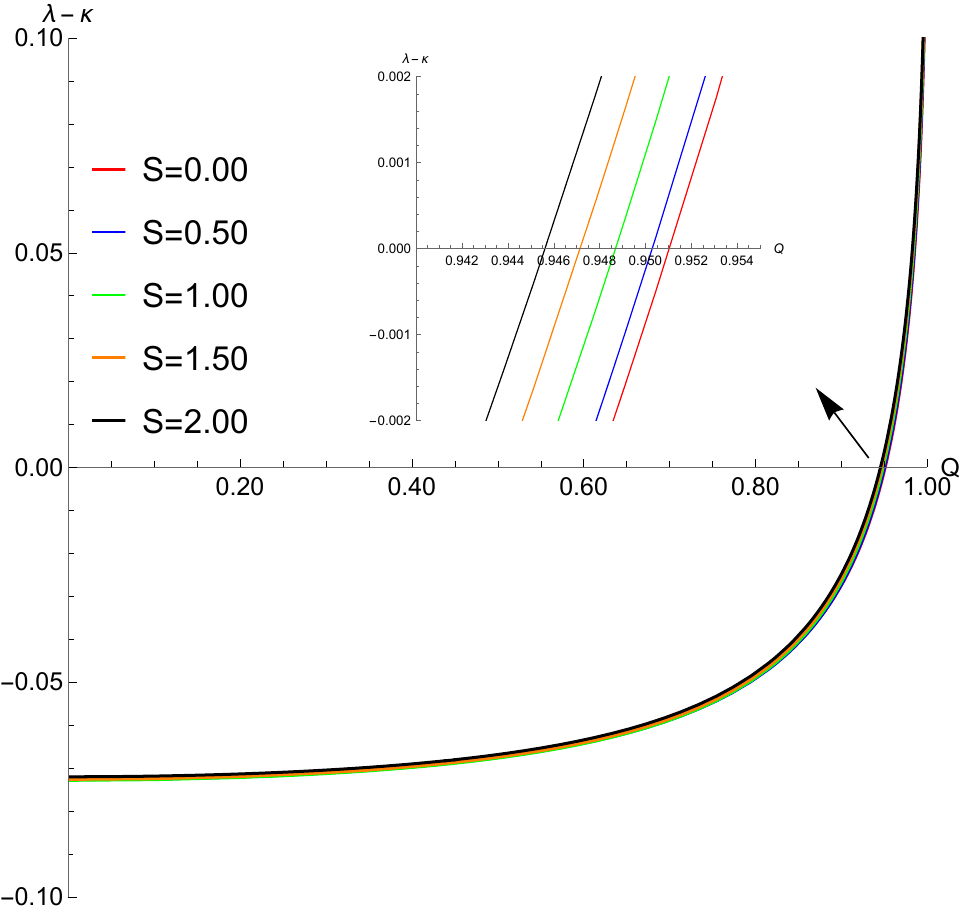}
		\subcaption{}
		\label{4f2-b}
	\end{minipage}
	\caption{Curves display the relationship between the deviation from the chaos bound and the BH's charge, with $L=10.00$ employed in the calculations. The left graph depicts the scenario wherein the spin direction is anti-aligned with the total angular momentum direction, while the right graph presents the case where they are aligned.}
	\label{4f2}
\end{figure}

We fix the angular momentum and then proceed to a discussion on the influence of the BH's charge on the exponent under diverse spin scenarios. In Figure \ref{4f2-a}, it is observed that when the spin is held constant, an increase in the charge causes the values of the exponent to approach and eventually exceed the surface gravity. The magnitude of the deviation from the gravity grows with the charge, indicating violations of the bound when the charge surpasses the critical thresholds. As the spin increases, the critical charge required to breach the bound progressively decreases. Consequently, the higher spin expand the range of the charges capable of violating the bound. When $Q=1.00$, the BH becomes extremal, and the violation is present regardless of the spin's value. Conversely, when $Q = 0$, the metric reduces to the Schwarzschild metric, and there is no violation for all spin configurations. This behavior is consistently replicated in Figure \ref{4f2-b}. When the spin direction aligns with the angular momentum direction, the effect of spin magnitude on the difference between the exponent and surface gravity is less pronounced than that of the charge. This indicates that both the charge and the spin simultaneously affect the violation, yet the influence of the former greater that of the latter. In the subsequent figures, the BH's charge is uniformly set to $Q=0.95$.

\begin{figure}[h]
	\begin{minipage}[t]{0.8\textwidth}
	\centering
	\includegraphics[width=8cm,height=6cm]{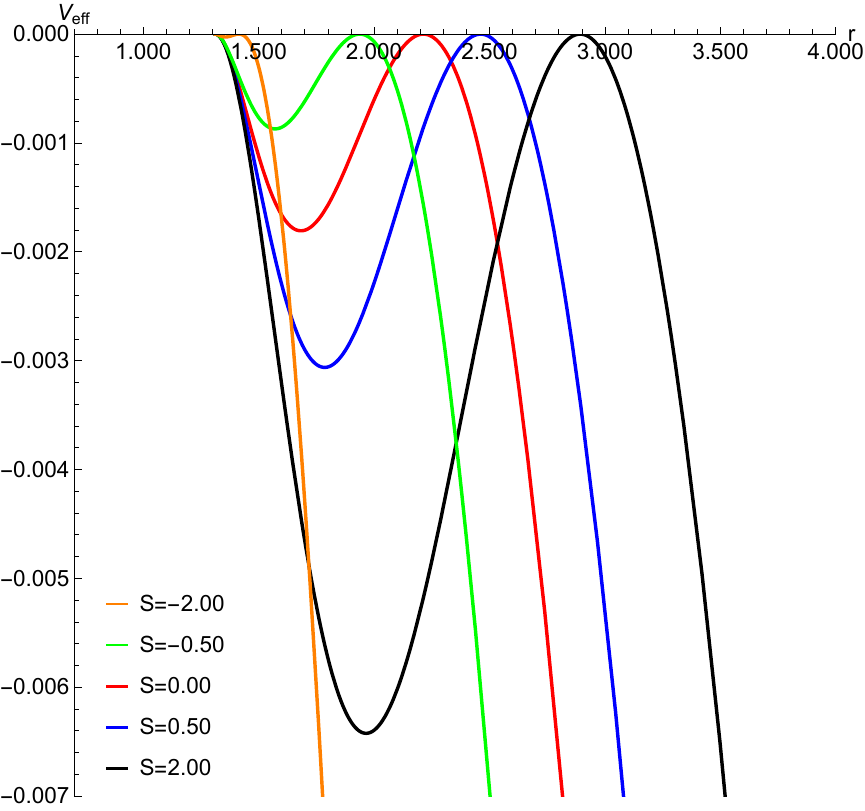}
\end{minipage}
\caption{Curves show the variation of the effective potential of the neutral particle with its radius, calculated with $L=10.00$.}
\label{4f1}
\end{figure}

We initiate our analysis by computing the effective potential and employing Figure \ref{4f1} to illustrate its variation with respect to the radial radius. A detailed observation of the figure reveals that an augmentation in the spin results in a shift to the right of the radial position corresponding to the maximum value of the effective potential. This, consequently, causes an expansion in the distance separating the equilibrium orbit from the event horizon. Therefore, when the spin direction is anti-aligned with the angular momentum direction, the position of the unstable equilibrium orbit is closer to the event horizon compared to the case without spin. Conversely, when the two directions are aligned, the position of the orbit is farther away from the horizon than in the spinless case.

\begin{figure}[h]
	\centering
	\begin{minipage}[t]{0.8\textwidth}
		\centering
		\includegraphics[width=8cm,height=6cm]{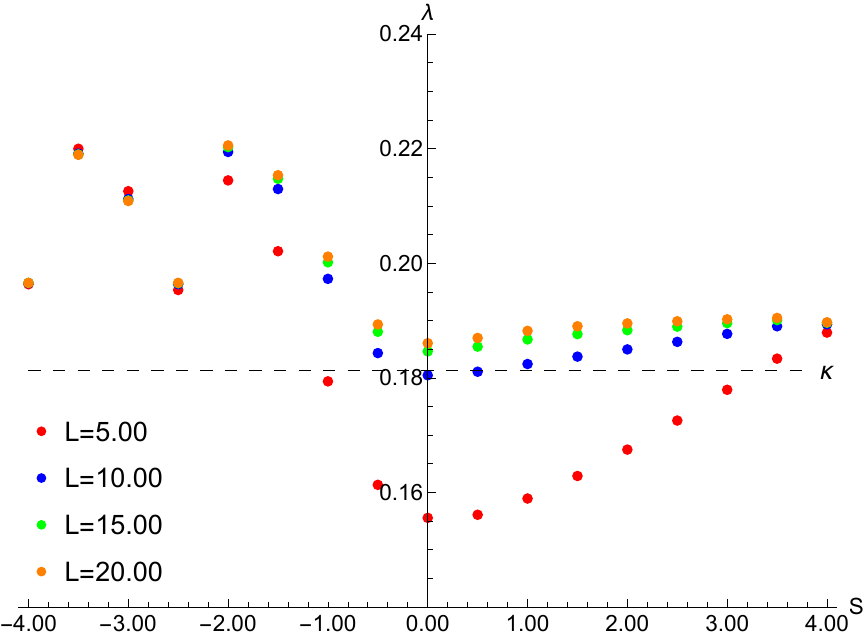}
		\subcaption{}
		\end{minipage}
\caption{The variation of the LE for the chaos of the neutral particle with respect to its spin. }
\label{4f3}
\end{figure}

Figure \ref{4f3} illustrates how the exponent varies with the spin under two scenarios: when the spin direction is aligned and anti-aligned with the angular momentum direction. In scenarios where their directions are aligned, for a given angular momentum, the value of the exponent rises as the spin increases, with these values corresponding to different angular momenta gradually converging. This suggests that the spin serves to enhance the chaotic behavior exhibited by the particle. For a relatively small angular momentum (e.g., $L=5.00$), the exponent changes relatively rapidly with an increase in the spin. The violation occurs only once the spin surpasses the specific threshold. Conversely, for a large angular momentum (e.g., $L=20.00$), the exponent changes more slowly with an increase in the spin, and even a zero spin can lead to the violation in this case. When the two directions are anti-aligned, it becomes evident that the exponent exhibits two maxima as the spin magnitude increases for a fixed angular momentum. When the angular momentum is not less than $15.00$, the violation emerges regardless of the spin's value. However, when the angular momentum is not greater than $10.00$, a violation manifests solely when the spin exceeds the threshold. This indicates that both the spin and angular momentum simultaneously influence the exponent, with the spin taking a dominant role for large spin magnitudes. Under the conditions of the identical spin magnitude and angular momentum, when the spin direction is anti-aligned with that of the angular momentum, the value of the exponent consistently exceeds that observed when the two directions are aligned. This indicates that the chaotic behavior of the particle is more pronounced when its spin direction is anti-aligned with the angular momentum direction compared to when they are aligned.

\begin{figure}[h]
	\centering
	\begin{minipage}[t]{0.48\textwidth}
		\centering
		\includegraphics[width=7cm,height=6cm]{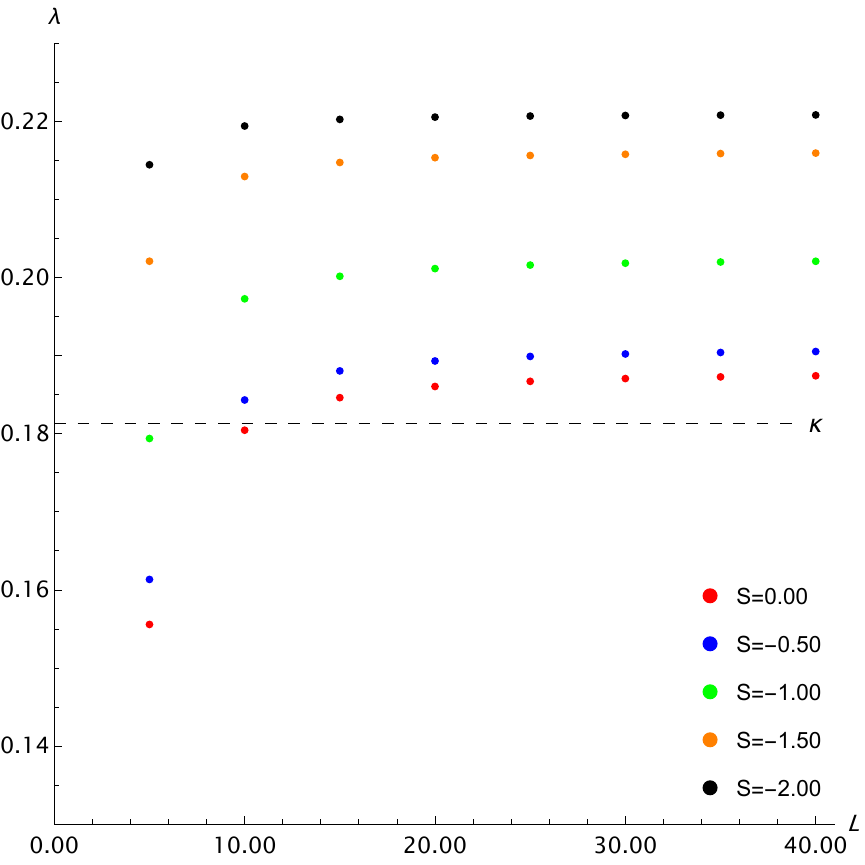}
		\subcaption{}
		\label{4f4-a}
	\end{minipage}
	\begin{minipage}[t]{0.48\textwidth}
		\centering
		\includegraphics[width=7cm,height=6cm]{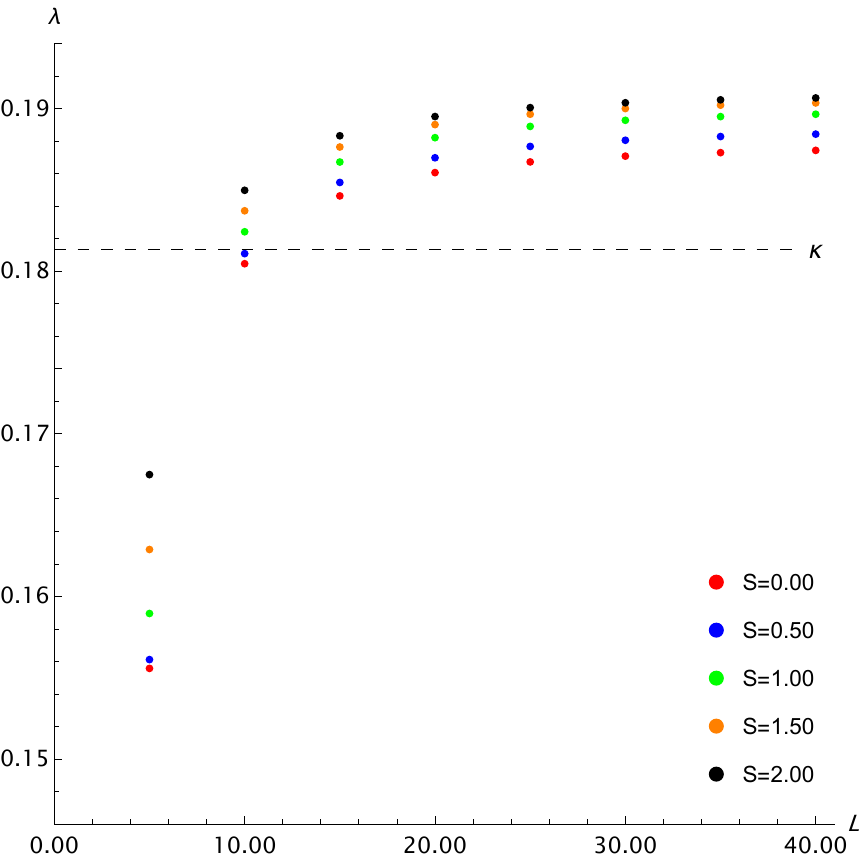}
		\subcaption{}
		\label{4f4-b}
	\end{minipage}
		\caption{The variation of the LE for the chaos of the neutral particle with respect to its total angular momentum.} 
	\label{4f4}
\end{figure}

The variation of the exponent against the angular momentum is plotted in Figure \ref{4f4}. It is observed that the exponent increases monotonically with the angular momentum for a fixed spin. This implies that an increase in the angular momentum intensifies the chaotic behavior of the particle. When the angular momentum is small, the exponent exhibits a relatively rapid variation with respect to the angular momentum. Conversely, as the angular momentum increases, the rate of change in the  exponent diminishes significantly. For a fixed spin magnitude, the exponent is significantly higher when the spin direction is anti-aligned with the angular momentum direction compared to the aligned case. Specifically, at $L=10.00$, the exponent for $S=-0.50$ exceeds that for $S=0.50$. The former surpasses the surface gravity, thereby violating the bound, while the latter remains below the surface gravity, preserving the bound. This indicates that the chaos is more pronounced when their  directions are anti-aligned. In this figure, except for the cases with $S=-1.50$ and $S=-2.00$, violations of the bound manifest when the angular momentum exceeds the specific critical thresholds.

\subsection{Charged particle}\label{sec3.3}

When the particle carries an electric charge, the electromagnetic force exerted on it by the BH affects the position of its unstable equilibrium orbit as well as the value of the exponent. In this regard, we carry out an investigation into the influences exerted by the charge, spin and total angular momentum of the particle on the exponent. We also use Eqs. (\ref{eq3.2}), (\ref{eq3.2.0}), (\ref{eq3.17}), (\ref{eq3.18}), (\ref{eq3.19}) and (\ref{eq3.26}) to calculate the exponent and generate plots. Based on the computation presented in the preceding subsection, we continue to set $Q=0.95$ in this analysis.

\begin{figure}[h]
	\begin{minipage}[t]{0.8\textwidth}
		\centering
		\includegraphics[width=8cm,height=6cm]{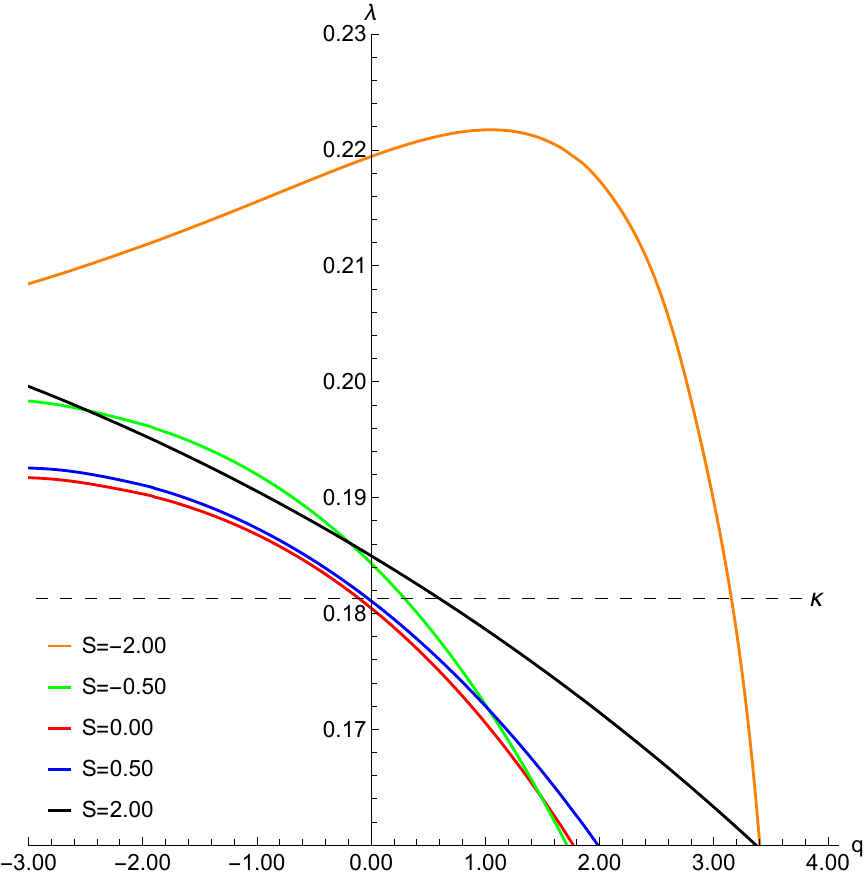}
	\end{minipage}
	\caption{Curves illustrate that the variation of the LE for the chaos of the charged particle with respect to its charge, with $L=10.00$ used in the calculations.}
	\label{5f2}
\end{figure}

We initially examine, through the variation of the particle's charge, whether the bound is violated in Figure \ref{5f2}. As shown in the figure, when the charge remains below a specific critical threshold, the exponent consistently exceeds the surface gravity, thereby violating the  bound. When the spin and angular momentum directions are aligned, it can be observed that, for a given charge, a larger spin magnitude corresponds to a higher value of the exponent. This situation also applies when the two directions are anti-aligned. This implies that an increase in the spin magnitude intensifies the chaotic behavior of the particle. Furthermore, it is shown here that both scenarios, where the charge-to-mass ratio of the particle is greater than $1.00$ or less than $1.00$, can lead to the violations. Given this, we order $q=0.70$ in the subsequent calculations. 

\begin{figure}[h]
	\begin{minipage}[t]{0.8\textwidth}
		\centering
		\includegraphics[width=8cm,height=6cm]{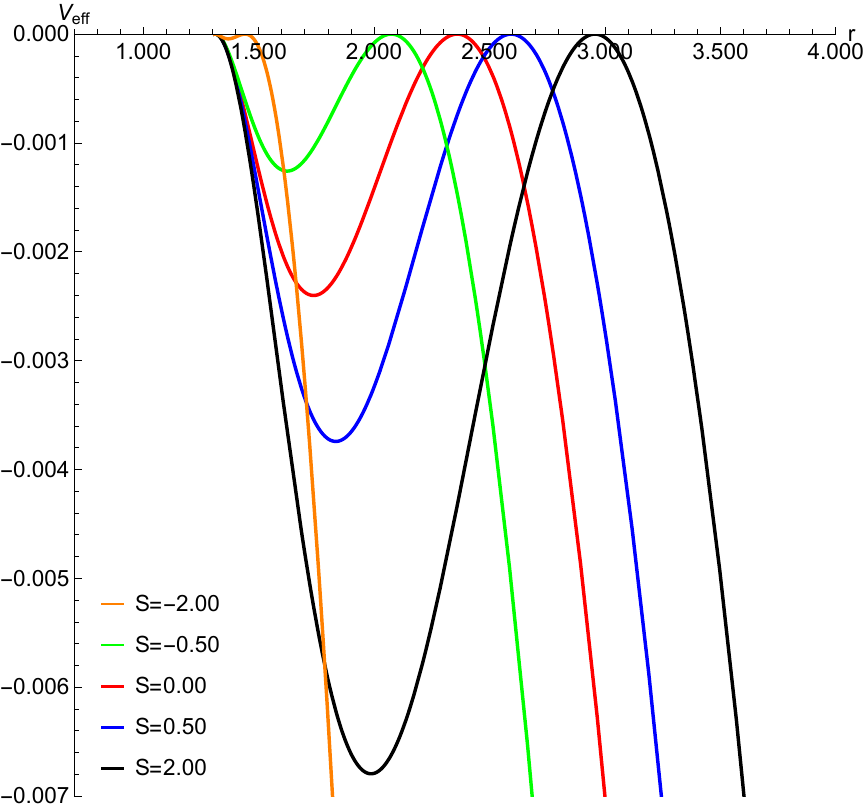}
	\end{minipage}
	\caption{Curves show the variation of the effective potential of the charged particle with its radius, calculated with $L=10.00$.}
	\label{5f1}
\end{figure}

The relationship between the effective potential and radius of the charged particle is plotted in Figure \ref{5f1}. This figure bears similarities to Figure \ref{4f1}, namely, the maxima of the potential all exhibit an increasing trend as the spin rises. Nevertheless, owing to the comparatively weak electromagnetic force, the discrepancy in this particular scenario isn't large. For a given spin, these maxima of the potential for the charged particle are larger than those for the neutral particle. Consequently, for the particle with the same spin, the unstable equilibrium orbit of the charged particle is farther from the event horizon compared to that of the neutral particle.

\begin{figure}[h]
	\begin{minipage}[t]{0.8\textwidth}
		\centering
		\includegraphics[width=8cm,height=6cm]{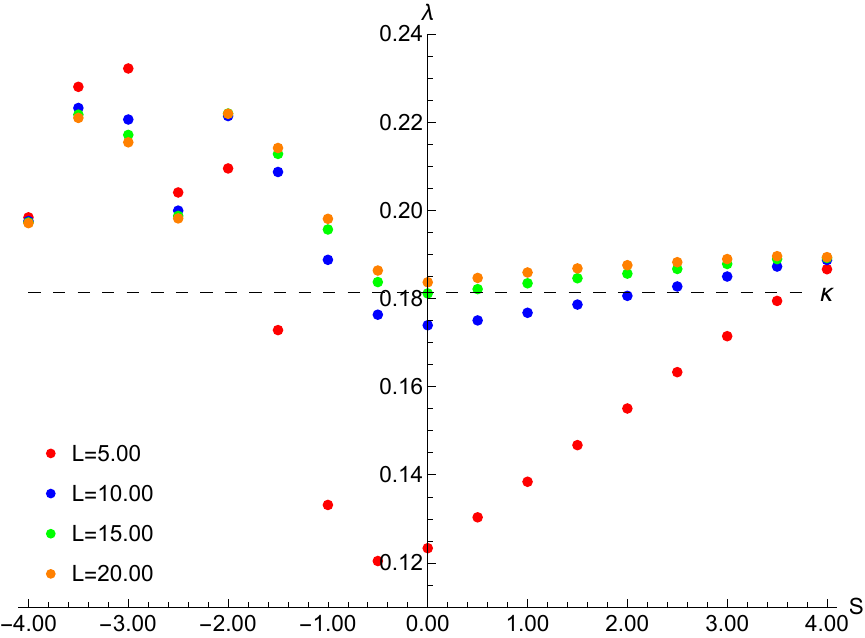}
	\end{minipage}
	\caption{The variation of the LE for the chaos of the charged particle with respect to its spin.} 
	\label{5f3}
\end{figure}

The variation of the exponent with respect to the spin is depicted in Figure \ref{5f3}. In this figure, two situations still persist: one in which the spin direction is aligned with the angular momentum direction, and the other in which they are anti-aligned. When the two directions are aligned, the trend of the exponent changing with the spin is similar to that in Figure \ref{4f3}. However, the threshold values of the angular momentum corresponding to the violation are not identical. For $L=15.00$, the bound for the spinless particle is not violated. When the two directions are anti-aligned, except for the case of $L = 5.00$, the variation of the exponent with respect to the spin magnitude is similar to that shown in Figure \ref{4f3}. We first elaborate on the situation associated with this angular momentum. In this scenario, the exponent has three minima. As the spin magnitude increases, it first decreases and then increases, presenting a minimum value smaller than the surface gravity when $S =-0.50$. The spin magnitude corresponding to the maximum value of the exponent has changed, showing a decrease compared to the spin magnitude in Figure \ref{4f3}. The possible cause of the aforementioned phenomenon might be that the electromagnetic force is not sufficiently strong. Consequently, when the electromagnetic force is relatively weak and the angular momentum is greater than $ 5.00$, this electromagnetic force cannot alter the trend of how the exponent varies with respect to the spin. Instead, it can only affect the magnitude of the exponent, which, in turn, influences the spin's threshold for violating the bound.

\begin{figure}[h]
	\centering
	\begin{minipage}[t]{0.48\textwidth}
		\centering
		\includegraphics[width=7cm,height=6cm]{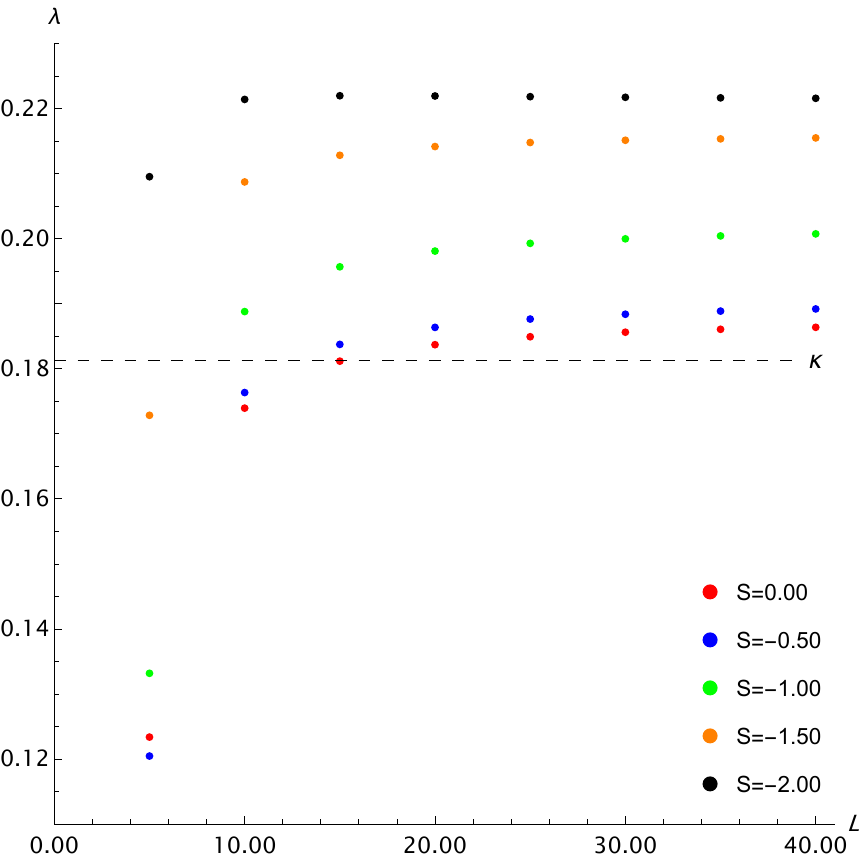}
		\subcaption{}
		\label{5f4-a}
	\end{minipage}
	\begin{minipage}[t]{0.48\textwidth}
		\centering
		\includegraphics[width=7cm,height=6cm]{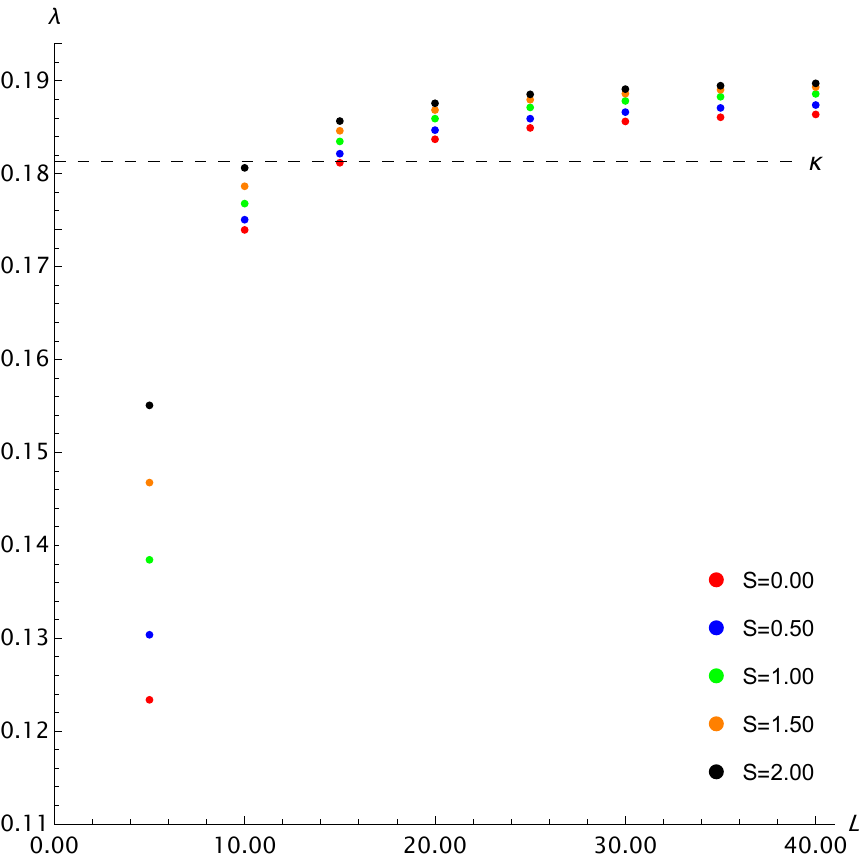}
		\subcaption{}
		\label{5f4-b}
	\end{minipage}
	\caption{The variation of the LE for the chaos of the charged particle with respect to its total angular momentum.} 
	\label{5f4}
\end{figure}

Figure \ref{5f4} illustrates the variation of the exponent with respect to the total angular momentum of the charged particle. When the spin direction is aligned with the angular momentum direction, for a given fixed spin magnitude, the exponent increases as the angular momentum rises. However, the angular momentum must surpass a specific threshold for the exponent's value to exceed the surface gravity, thereby resulting in a violation of the bound. Consequently, an increase in the angular momentum intensifies the chaotic behavior of the particle. The values of the exponent under specific angular momentum and spin conditions are generally smaller than those depicted in Figure \ref{4f4} for the corresponding cases. When the two directions are anti-aligned, for a fixed spin magnitude, the exponent still exhibits an increasing trend with the rise in the angular momentum. Nevertheless, at the spin magnitude of $2.00$, a violation occurs irrespective of the angular momentum values. For other spin magnitudes, the violations only manifest when the angular momentum exceeds certain thresholds.

\section{Conclusions and discussions}\label{sec4}

In this work, we explored the motions of the spinning particles around the RN BH, and examined the chaos bound. Taking into account the two cases where the spin direction of the particle is aligned and anti-aligned with its total angular momentum direction, we found that the LEs can surpass the surface gravity of this BH under the specific conditions. Therefore, the phenomenon of the chaos bound violation was found in the spinor field. 

For the neutral particle, we initially fixed its total angular momentum and analyzed the influence of the BH's charge on the bound. It was discovered that the violation occurs only when the charge exceeds its threshold values, and the different spins correspond to the distinct charge thresholds. Based on these thresholds, we further fixed the charge and examined the impacts of the spin and angular momentum on the exponent. For a given angular momentum, when the spin direction aligns with the angular momentum direction, the exponent increases with the rising spin. Conversely, when their directions are opposite, the exponent displays two maxima as the spin magnitude increases. When the particle is charged, we took into account the influence of the electromagnetic force. It was observed that there are certain similarities in the variation of the exponent with the spin between the charged and neutral particles (except for the case where the angular momentum is $5.00$). This indicates that the electromagnetic force cannot change the trend of how the exponent varies with the spin and angular momentum. However, it can alter the values of the exponent and result in the violations. The likely reason for this phenomenon is that the electromagnetic force is not sufficiently strong. If both the BH's and particle's charges are large enough simultaneously, the trend of the exponent's variation may change.

There are two explanations for the violation. The first explanation admits the violation, and its proponents posit that this violation may be associated with the thermodynamic stability of BHs. However, the subsequent research has demonstrated that when the BH undergoes the phase transition, the violation occurs in the BH branch within its stable region \cite{LGD}. The second explanation asserts that there is no real violation. Instead, such violations can be avoided by introducing the effective temperature \cite{HT5} or by modifying the formulation of the bound \cite{LTW1}. On the other hand, the chaos bound was originally proposed within the framework of thermal quantum subsystems, whereas our calculations of the LEs were carried out in the classical system. In the calculations, we adopted the particle's mass of unity without accounting for the backreaction of the particle on the background spacetime. This backreaction has the potential to indirectly influence the exponents by modifying the gravitational field of the BH.  Therefore, to determine whether there is a violation for the bound, it is essential to incorporate these factors into the analysis.

\end{document}